\RequirePackage{fix-cm}

\documentclass[smallextended]{svjour3}  

\smartqed  

\usepackage{url}
\usepackage{cite}
\usepackage{wrapfig}
\usepackage{graphics}
\usepackage{picins,graphicx}
\usepackage[english]{babel}
\usepackage[leqno]{amsmath}
\usepackage{amssymb}
\usepackage[mathscr]{eucal}
\usepackage{amstext}
\usepackage{makeidx}
\usepackage{hyperref}
\usepackage{pdfsync}

\journalname{myjournal}

\date{Received: date / Accepted: date}

\newcommand{\beq}{\begin{equation}}
\newcommand{\eeq}{\end{equation}}
\newcommand{\no}{\noindent} 
 
\newcommand{\bex}{\begin{ex}} 
\newcommand{\eex}{\end{ex}} 
\newcommand{\bese}{\begin{esempio}} 
\newcommand{\eese}{\end{esempio}} 
\newcommand{\bpro}{\begin{proposition}} 
\newcommand{\epro}{\end{proposition}} 
 
\newcommand{\bthe}{\begin{theorem}} 
\newcommand{\ethe}{\end{theorem}} 
\newcommand{\bnote}{\begin{notation}} 
\newcommand{\enote}{\end{notation}} 
\newcommand{\bdefi}{\begin{definition}} 
\newcommand{\edefi}{\end{definition}} 
\newcommand{\bc}{\begin{center}} 
\newcommand{\ec}{\end{center}}

\begin{document}

\title{On  exponentially shaped Josephson junctions}

\author{Monica De Angelis }
\institute{M.De Angelis \at
             University of Naples Federico II, Faculty of Engineering. Dept. of Math. and Appl. \\ Via Claudio 21,  80125 Naples,  Italy \\
              \email{modeange@unina.it }}
\maketitle

\begin{abstract}
The paper deals with a third order semilinear equation which  characterizes exponentially shaped Josephson junctions in superconductivity. The initial-boundary  problem  with Dirichlet conditions is  analyzed. When the source term $ F  $  is a linear function, the problem is explicitly solved by means of a Fourier series
with properties of rapid convergence. When $ F $   is nonlinear, appropriate estimates of this 
series allow to deduce a priori estimates, continuous dependence and  asymptotic
behaviour of the solution.
\keywords{ Superconductivity\and Josephson junction\and  Partial  differential equations \and Fundamental solution.  }
\PACS{85.25\and 74.81 Fa}
\subclass{ 35K35\and 35E05  }

\end{abstract}

\section{Introduction }
\label{intro}

We refer to the semilinear equation

 \beq                                   \label{11}
 {\cal L}_ \varepsilon u=
F(x,t,u)
 \eeq

\noindent
where $\ {\cal L}_ \varepsilon $ is the third - order parabolic operator:

 \beq                                   \label{12}
 {\cal L}_ \varepsilon =(\partial_{xx}-\lambda \partial_x)(\varepsilon
\partial_t+1 ) - \partial_t(\partial_t+\alpha).
 \eeq

The equation (\ref{11}) characterizes the evolution of several
dissipative models such as the
motions of viscoelastic fluids or solids\cite{bcf,dr1,jp,r}; the sound propagation in
viscous gases \cite{l}; the heat conduction at low temperature\cite{jcl,mps} and  the propagation of localized magnetohydrodinamic models in plasma physics \cite{sbes}. Moreover, it can  also be referred  to reaction diffusion systems \cite{acscott}. 

As  example of  perturbed model of the phase evolution,  we
will consider  the  non linear phenomenon concerning the
Josephson effects in superconductivity.

More precisely, if  $\varphi=\varphi(x,t)$ is the phase difference in a rectangular junction and  $\gamma$ is the normalized current bias; when
  $  \lambda\,=\,0 \,\,$\,and $F=sin\varphi -\gamma$, the equation (\ref{11}) gives the well-known  perturbed Sine-Gordon equation (PSGE)  \cite{bp}:

 \beq           \label{13}
  \varepsilon \varphi_{xxt}+\varphi_{xx}-\varphi_{tt}-\alpha\varphi_{t}=\sin
  \varphi-\gamma.
 \eeq

The terms $\,\varepsilon \varphi_{xxt}\,$ and $\,\alpha\,\varphi_t\,$ characterize the
 dissipative normal electron current flow respectively along and across the junction. They represent the
{\em perturbations} with respect to the classic Sine Gordon
 equation \cite{bp,scott}.  When the
 surface resistance is negligible, then $\varepsilon \,(<1)$ is
 vanishing and a singular perturbation problem for the equation
 (\ref{13}) could  appear \cite{for}.
 As for the coefficient $ \alpha$ of (\ref{13}), it depends on
 the shunt conductance \cite{csr} and generally one has $a< 1$
 \cite{j005,lom1,pag1}. However, if the resistance of the junction is so small as to
 short completely the
 capacitance, the case  $a>1$ arises \cite{bp,tin}.
 
 More recently, the case of the exponentially shaped Josephson junction (ESJJ) has been   considered. The evolution of the phase inside this junction is described  by the third order equation:

 \beq            \label{14}
  \varepsilon \varphi_{xxt}+\varphi_{xx}-\varphi_{tt} - \varepsilon\lambda \varphi_{xt} -\lambda  \varphi_x- a\varphi_{t}=\sin
  \varphi-\gamma
 \eeq

 \no where $ \lambda$ is a positive constant generally less than one \cite{bcs96,bcs00} while the terms  $ \,\,\lambda \,\varphi_x $ and $\,\lambda \, \varepsilon \,\varphi_{xt} \,$ represent the current due to the tapering. In particular $ \lambda\varphi _{x} $ correspond to a geometrical force driving the fluxons from the wide edge to the narrow edge. \cite{bcs00,cmc02}.

 \no According to  recent literature \cite{bcs96,bss,cmc02,j05,j005,ssb04}, an  exponentially shaped Josephson junction  provides several advantages with respect to a rectangular junction. For instance in \cite{bcs96} it has been proved that in an ESJJ  it is possible to obtain  a voltage which is not  chaotic anymore, but rather periodic  excluding, in this way,   some among the possible causes of large spectral width. It is also proved that  the problem of trapped flux can be avoided. Moreover, some devices as SQUIDs were built with exponentially tapered loop areas.\cite{cwa08}

The analysis of many initial -
 boundary problems related to the PSGE (\ref{13}) has been
discussed in a lot of papers. In particular in\cite {mda01}, to deduce an exhaustive
asymptotic analysis, the 
Green function of the linear operator 
\beq                                   \label{15}
 {\cal L}=\partial_{xx}(\varepsilon
\partial_t+c^2 ) - \partial_t(\partial_t+\alpha)
\eeq

 \no has been determined by Fourier series. By means of its properties an exponential decrease of both  linear and non linear solutions is deduced.

The aim of this paper is the analysis of the Dirichlet boundary value problem related to the   equation (\ref{14}).

\no
The Green function$ \,G\, $ of the linear strip problem is determined by Fourier series and properties of rapid convergence are established. So, when the source term $ F $ is a linear function, then the explicit solution is obtained and an exponential decrease of the solution is deduced.

When $ F $ is nonlinear, the problem is reduced to an integral equation with kernel $ G $ and  an appropriate analysis implies results on the existence and uniqueness of the solution. Moreover, by means of suitable properties of $ G ,  $  a priori estimates, continuous dependence upon the data  and   asymptotic behaviour of the solution are achieved, too.

\setcounter{section}{1}

 \section{Statement of the problem and properties of the Green function}

\label{sec:2}

Let $l , T  $ be arbitrary positive constants and let

\vspace{3mm}

$\ \ \ \ \ \ \ \ \ \ \ \  \ \ \ \ \ \ \   \Omega_T =\{(x,t) : 0 < x <
l, \  \ 0 < t \leq T \}$.

\vspace{3mm}\noindent
 The boundary value  problem related to equation (\ref{11}) is the following:

  \beq          \label{21}
  \left \{
   \begin{array}{ll}
    & (\partial_{xx} \, - \,\lambda\,  \partial _x\,)\,\,(\varepsilon
u _{t}+ u) - \partial_t(u_{t}+\alpha\,u)\,=F(x,t,u),\ \  \
       (x,t)\in \Omega_T,\vspace{2mm}\\
   & u(x,0)=h_0(x), \  \    u_t(x,0)=h_1(x), \  \ x\in [0,l],\vspace{2mm}  \\
    & u(0,t)=0, \  \ u(l,t)=0, \  \ 0<t \leq T.
   \end{array}
  \right.
 \eeq

By Fourier method it is possible to determine the Green function of the linear  operator ${\cal {L}_\varepsilon }$.
So,  let

\beq                                             \label{22}
\gamma_n=\frac{n\pi}{l},\  \quad  b_n=(\gamma_n^2\,+ \lambda^2/4\,), \quad  g_n=\frac{1}{2} (\alpha\,+\,\varepsilon\,b_n\,),\  
\  \omega_n=\sqrt{g_n^2- \,b_n\,}
\eeq

\noindent and

\beq                                \label{23}
G_n(t)= \,\,\frac{1}{\omega_n}\,\,e^{-g_n\,t}\,\,
sinh(\omega_nt),
\eeq

\no by standard techniques, the Green function  can be
given the form:

\beq                                                \label{24}
G(x,t,\xi)=\frac{2}{l}\,\, e^{\frac{\lambda\,}{2\,}\,x}\,\,\sum_{n=1}^{\infty}\,
G_n(t) \  \  sin\gamma_n\xi\  \ sin\gamma_nx.
\eeq

\noindent
This series is endowed  of rapid convergence and it is  exponentially vanishing as t tends to infinity. In fact, if we denote by

\beq    \label{25}
\, a_\lambda = \alpha \, + \varepsilon \, \lambda^2/4    
\eeq
\no  and 

\beq                      \label{26}
p_\lambda= \frac{\pi^2}{\varepsilon\,\pi^2 \,+\,a_\lambda\,l^2},  \ \ \  q_\lambda=  \frac{\,a_\lambda\,+\,\varepsilon(\pi/l)^2}{2}, \ \ \ \delta\equiv min(p_\lambda, q_\lambda),
\eeq

\no the following theorem holds:

\begin{theorem}   \label{t21} 
Whatever the constants $\alpha, \varepsilon, \lambda $
 may be in $\Re^+$, the function $G(x,\xi,t)$ defined in
(\ref{24})  and all its time derivatives are continuous functions in
$\Omega_T$ and it results:

\beq                                        \label{27}
|G(x,\xi,t)|\,\leq \,M  \,e^{-\delta t}, \ \ \, \ \ \ \
|\frac{\partial ^j G}{\partial t^j}|\, \leq \,N_j \,  e^{-\delta
t},
\\ \ \ \ \ \ \, \ \ j \in {\sf N}
\eeq

\noindent where  $M, N_j $ are constants depending on  $\alpha, \lambda, \varepsilon$.
\end{theorem}

{\bf Proof}. 
Physical problems lead to consider  $\alpha \,\varepsilon \,<1$ and denoting by

\beq                              \label{28}
  N_{1,2}^\lambda = \frac{l}{2\,\varepsilon^2\pi}\biggl[ \,4\, \bigl(1\mp \sqrt{1-\alpha \,\varepsilon} \, \bigr)^2 - {\varepsilon^2\, \lambda^2}\biggr]^{1/2},
\eeq

\no  let us assume that  $  N_{1,2}^\lambda\,>1 $. So, let  $ k$ be a positive constant
less than one and let  
 $\bar   N_{1,2} , N_k^\lambda $  be the lowest integers such that

\beq
\left \{
   \begin{array}{ll}
    & \bar N_1 <  N_1^\lambda, \ \qquad   \bar N_2 \ > N_2^\lambda; \\ \vspace{2mm}\\
   & N_k^\lambda \ >\,\,\dfrac{l}{2\,\varepsilon^2\pi\, {k}}\,\biggl [ \,4\, \bigl(1\mp\,\sqrt{1-\alpha \,k\,\varepsilon} \, \bigr)^2 - {\varepsilon^2\, k\,\lambda^2}\biggr]^{1/2}.
   \end{array}
  \right.
\eeq

We start   analysing  the hyperbolic terms  when
$n\geq \bar N_2. $   Letting

\beq                                      \label{210}
X_n=\dfrac{b_n}{g_n^2}\,<1\,\,\qquad \qquad and \qquad \qquad\varphi_n=g_n(-1+\sqrt{1-X_n}),
\eeq

 \no it is possible to prove that $ \varphi_n\,\,\leq\,- \dfrac{\gamma_n^2}{2\, g_n}. \,\,$ So it results:

\beq                                       \label{211}
e^{-t(g_n-\omega_n)}\leq \ \ e^{-p_\lambda \,t}.
\eeq

\no Furthermore, it is easily verified that  for  all $n \, \geq N^\lambda_k \,(\,\geq \bar N_2) $  it results $ \dfrac{b_n}{g_n^2}\,\leq \, k $ and hence one has:

\beq                          \label{212}
\omega_n \, \geq g_n \ (1-k)^{1/2} \,\geq\, n^2 \, \frac{2 l^2}{\varepsilon\, \pi^2} \,\,(1-k)^{1/2}.
\eeq

Other  terms  can be treated similarly. For instance, as for  circular terms, it can be  proved that  $ e^{-\,g_n\,t}\leq \ \ e^{-q_\lambda \,t}$.

\no In  consequence  estimate
$(\ref{27})_1$ holds $ \forall \, n \geq 1. $

 \vspace{3mm} As for $(\ref{27})_2,$ one has :

\beq                       \label{213}
g_n\,- \omega_n\,=\,\,\frac{b_n}{g_n + \omega_n\,}\, \leq \, \frac{2}{\varepsilon\,}+\frac{\lambda^2}{4\,\,\varepsilon\,q_\lambda}\qquad \qquad  \forall \,n \geq 1
\eeq

\no and by means of standard computations, $(\ref{27})_2$ can be  deduced, too.

It may be similarly proved that  the theorem holds also when    $ \alpha\,\varepsilon \geq\,1  $  or when the conditions $N_{1,2}^\lambda\, > 1 $  do not hold.

 Finally we notice that  when $N_{1,2}^\lambda\,$ are integers,  the constants $M$ and $N_j$ in (\ref{27}) could depend on t.

\vspace{3mm}As for the  x-derivatives of Fourier series like
(\ref{24}), attention is needed towards convergence problems. For this, we
will consider x-differentations  of the operator
$(\varepsilon\partial_t+1)G$       instead of   $G $ and   $G_t$.

\bthe \label{t22} 
Whatever $\alpha , \varepsilon, \lambda $  may be
, the function   $ G(x,\xi,t)$ defined in (\ref{24}) is such that:

\beq                            \label{214}
 |\partial_{x}^{(i)}\,\,(\varepsilon\, G_t\, +\, G )|\\ \,\leq A_i \, \,
\, e^{-\delta t},\qquad (i=0,1,2) 
\eeq

\noindent
 where $ \delta  $ is defined in (\ref{26}) and $A_i\,\,\,(i=0,1,2)$   are constants depending  on $a, \varepsilon ,\lambda $.
\ethe

{\bf Proof}.
As for the
hyperbolic terms in $G, $ it  results:

\beq                                    \label{215}
 \varepsilon  G_t+G=e^{\frac{\lambda\,x\,}{2\,}}\,\,\sum_{n=1}^{\infty} \,\,\frac{e^{-g_nt}}{l\,\,\omega_n} \{[1-\varepsilon( g_n-\omega_n)]
e^{\omega_nt}-[1-\varepsilon(g_n+\omega_n)]e^{-\omega_nt}\}.
\eeq

\noindent where according to (\ref{210}), it results: 

\[ 1-\varepsilon ( g_n-\omega_n) \, = \, 1\,+\varepsilon\, \varphi_n.
\]
\no So, by means of Taylor's formula, one has:

\beq                         \label{216}
1-\varepsilon \, ( g_n-\omega_n) \,\,= 1- 
\frac{\varepsilon}{2}\,
g_n\,X_n\,-\,\dfrac{\varepsilon }{8Ã¢â‚¬Â¢} g_n \, X_n^2    -\frac{3}{16} \, \varepsilon \, g_n \,\int_0^{X_n}\frac{(X_n-y)^2}{(1-y)^{5/2}} \,\  dy.
\eeq

\noindent
Besides, it is possible to prove that $ \forall n\geq  1$ one has:

\beq                         \label{217}
X_n < \, \frac{c^2}{n^2},\ \  \, \, \ \ \ \ \ (c =
l\,\,\sqrt{4+\lambda^2}/\,\,\varepsilon\,\,\pi)
\eeq

\noindent
and for all   $n\geq \, c(1+c_1)  (c_1>0)$ it results:

\beq                       \label{218}
\int_0^{X_n}\frac{(X_n-y)^2}{(1-y)^{5/2}} \,\  dy\, \leq \frac{2}{3} \ X_n^2 \,
\biggl[   \dfrac{(c_1+1)^3}{c_1(c_1+2)]^{3/2}}-1\biggr].
\eeq

\noindent
So, taking into account that

\[
\frac{\varepsilon}{2}\,
g_n\,X_n\, = \, 1- \dfrac{\alpha}{a_\lambda \,+ \varepsilon \, \gamma _n ^2Ã¢â‚¬Â¢} \]
\[\,\dfrac{\varepsilon }{8Ã¢â‚¬Â¢} g_n \, X_n^2 =\, \dfrac{\varepsilon \, b_n ^2Ã¢â‚¬Â¢}{(a_\lambda \,+ \varepsilon \, \gamma _n ^2)^3Ã¢â‚¬Â¢}; \]

\no there exists  a positive constant $ k_1 $ such that :

\beq                         \label{219}
|1+\varepsilon \ \varphi_n | \ \, \leq \frac{1}{n^2} \
\bigl(\,\dfrac{\alpha l^2}{\varepsilon\pi^2Ã¢â‚¬Â¢}\,+\dfrac{k_1}{n^2Ã¢â‚¬Â¢}\,\bigr).
\eeq

\noindent

\noindent
Estimates of theorem \ref{t21} together with (\ref{219}) show that the series terms related to the operator $ \varepsilon \,G_t + G\, $ have order at
least  of $n^{-4}$. So it can be differentiated term by term with respect
to x and the estimate (\ref{214}) can be deduced.

\vspace{2mm}
As solution of the equation
${\cal L}_\varepsilon v
 = 0$  we will mean a continuous function
$v(x,t)$ which has continuous the derivatives $v_t, v_{tt}, \partial_{x}(\varepsilon
v_t+v),  \,\,\partial_{xx}(\varepsilon
v_t+v) $  and these derivatives verify the equation.

\noindent
So,
 we are able to prove the following theorem:

\bthe \label{t23} 
 The
function  $G(x,t)$ defined in (\ref{24}) is a
solution of the equation

\beq                      \label{220}
{\cal L}_\varepsilon G \, =(\partial_{xx}\, -\, \lambda \, \partial_x)(\varepsilon
G _{t}+ G) - \partial_t(G_{t}+\alpha\,G)=0.
\eeq
\ethe

{\bf Proof}. The uniform convergence proved in theorems \ref{t21}-\ref{t22}
allows to deduce that:

\beq                      \label{221}
(\partial_t\,+\alpha)\frac{\partial G}{\partial t}=
\frac{2 e^{\frac{\lambda\,x\,}{2\,}}}{l} \,\sum_{n=1}^{\infty}\biggl\{\bigl[ b_n\bigl(\varepsilon  g_n -1\bigr)G_n
-\varepsilon\, b_n
e^{-g_nt}\cosh\omega_nt\bigr]\sin\gamma_n\xi
\sin
\gamma_nx\,\biggr\},
\eeq

\beq                           \label{222}
  \partial_{x}(\varepsilon\partial_t+1)G=\frac{2}{l}e^{\frac{\lambda\,x\,}{2\,}}\sum_{n=1}^{\infty}\biggl\{\biggl[\,\,\frac{\lambda }{2}\,\, \bigl(1- \varepsilon\,\,g_n\,\bigr)\,\,G_n  +
\eeq  
  \[+\,\,\frac{\lambda  \varepsilon }{2}
e^{-g_nt} \cosh\omega_nt\biggr]\sin\gamma_n\xi
\sin
\gamma_nx +\]
\[
\,\,+\bigl[\,G_n \,\, \gamma_n \,(1-\varepsilon\, g_n )+  \varepsilon \gamma_n e^{-g_nt} \cosh\omega_nt\bigr]\sin\gamma_n\xi
\cos
\gamma_nx\biggr\}. 
\]

\noindent
Moreover, beeing 

\beq                           \label{223}
  \partial_{xx}(\varepsilon\partial_t+1)G=\frac{2}{l}e^{\frac{\lambda\,x\,}{2\,}}\sum_{n=1}^{\infty}\biggl\{ \bigl[\bigl( \varepsilon\,\,g_n-\,\,1\bigr)\,\,\bigl(b_n - \frac{\lambda^2}{2}\bigr) \,\,G_n+
\eeq

  \[ \bigl(-\frac{\lambda ^2 \varepsilon }{4}+\varepsilon\gamma^2_n\bigr)
e^{-g_nt} \cosh\omega_nt\bigr]\sin\gamma_n\xi
\sin
\gamma_nx +
\]

\[+\bigl[ \, G_n \lambda \gamma_n (1- \varepsilon g_n) + \varepsilon \lambda  \gamma_n e^{-g_nt}\cosh\omega _n t\bigr]\sin\gamma_n\xi
\cos
\gamma_nx  \biggr\},
\]

\noindent
(\ref{220}) can be deduced.

 \section{Properties of the convolution}

To achieve the  solution of the strip  problem (\ref{21}), the  convolution of  the function G with the data must be analysed. For this,  let  $h(x) $ be a continuous function on $(0,l)$ and let:

\beq                             \label{31}
u_h(x,t)\ \ =
\int_{0}^{l} h(\xi) \, G(x,\xi,t)\ \ d\xi
\eeq

\beq                             \label{32}
u_h^*(x,t)\ \ =
(\partial_t+\alpha + \varepsilon \lambda \partial_x-\varepsilon\partial_{xx})u_h(x,t).
\eeq

\no  The following theorems hold:

\bthe   \label{t31} 
If the data $h(x)$ is a $C^1(0,l)$ function,
 then
$u_h$ defined by (\ref{31}) is a solution
  of the
equation ${\cal L}_\varepsilon =0$ and it results:

\beq                            \label{33}
\lim_{t \rightarrow 0}u_h(x,t)=0 \ \ \ \ \ \lim_{t \rightarrow 0} \partial_t
u_h(x,t)=h(x),
\eeq

\no uniformly for all $ x \in [0,l]$. 
\ethe

\vspace{3mm}{\bf Proof.} The absolute convergence  of $ u_h $  with its partial derivatives  is proved by means of theorems \ref{t21} and \ref{t22}  and  continuity of  function $ h(x). $ So, since (\ref{220}) $\,\,{\cal L}_\varepsilon   u_h=0\,\, $ is verified, while theorem \ref{t21} and hypotheses on $ h(x) $ imply $(\ref{33})_1$.   

More beeing:

\beq                            \label{34}
\dfrac{\partial G}{\partial \,t}\,=-\frac{2}{\pi} \ \ \frac{\partial}{\partial\xi} \ \ \sum_{n=1}^{\infty}
\ \ \dfrac{\partial G_n}{\partial \,tÃ¢â‚¬Â¢} \,\, \frac{\cos \gamma_n\xi}{n}\,\,\sin
\gamma_n x
\eeq
\no and 

\beq                            \label{35}
\partial_t u_h= -\frac{2}{\pi}
\sum_{n=1}^{\infty}
\ \ \frac{\partial G_n}{\partial t} \ \ [h(\xi)\cos \gamma_n\xi \
]^{l}_0 \ \ \dfrac{\sin\gamma_n x }{n}\  \eeq

\[+\frac{2}{\pi} \int_{0}^{l}\sum_{n=1}^{\infty}
\ \ \dfrac{\partial G _nÃ¢â‚¬Â¢}{\partial tÃ¢â‚¬Â¢} \,h^{'}(\xi) \, \frac{\cos \gamma_n\xi}{n} \,\ d\xi\,\,\sin
\gamma_n x, \]

\no  denoting by  $\eta(x)$   the Heaviside function, it results: 

\beq                            \label{36}
 \lim_{t \rightarrow 0}\partial_t\,u_h=
\frac{x}{l}\bigl[\,h(l)-h(0)\bigr]+h(0)- \int_{0}^{l} h^{'}(\xi)
\bigl[\,\,\eta(\xi-x)+\frac{x}{l}-1\bigl]
d\xi  =h(x).
\eeq

\bthe   \label{t32} 
 Let  $h(x) $ be a $ C^3(0,l)$  function  such that $h^{(i)}(0)=h^{(i)}(l)=0
  \,\,(i=1,2,3)$. 
Then  $u_h^*$  defined in (\ref{32})  is a solution
 of the
equation ${\cal L}_\varepsilon =0$  and it results:

\beq                            \label{37}
\lim_{t \rightarrow 0}u_h^*(x,t)=h \ \ \ \ \ \lim_{t \rightarrow 0} \partial_t
u_h^*(x,t)=0
\eeq

\noindent
uniformly for all $x\in (0,l)$.

\ethe

{\bf Proof.}  Properties of $ h(x) $ assure that:

\beq                               \label{38}
( \, \lambda \,\partial _x\,-\, \partial_{xx})\,u_h\,(x,t)\, = \,\eeq
\
\[=\,-\,\,\frac{2}{l} ( \, \lambda \partial _x- \partial_{xx})\,\,e^{\frac{\lambda\,x\,}{2\,}} \,
\sum_{n=1}^{\infty} G_n(t) \int_{0}^{l} h^{''}(\xi) \ \ \frac{\sin\gamma_n\xi}{\gamma_n^2}\ \ d\xi \,\, \sin \gamma_n x =\] 

\[=\,-\,\,\frac{2}{l} \,\,e^{\frac{\lambda\,x\,}{2\,}} \,
\sum_{n=1}^{\infty}\biggl[ G_n(t)\, \bigl( \gamma_n^2+\dfrac{\lambda^2Ã¢â‚¬Â¢}{4Ã¢â‚¬Â¢}\bigr)\ \int_{0}^{l} h^{''}(\xi) \ \ \frac{\sin\gamma_n\xi}{\gamma_n^2}\ \ d\xi \,sin \gamma_n x\,\biggr] =
\]

\[=  -\,\,u_{h^{''}}(x,t)+  \frac{\lambda^2}{4} \,\,u_{h}(x,t)
\]

\vspace{3mm} \no So,  since theorem \ref{t23},$\,\,{\cal L}_\varepsilon   u^*_h=0\,\, $ is verified. Moreover, beeing:

\beq                                    \label{39}
\partial_t u_h^* =   (\partial_{xx}- \lambda \partial _x) u_h
\eeq

\noindent     (\ref{38}) implies $(\ref{37})_2 $, too. 
Finally,  owing to (\ref{33}) and (\ref{38}) , one obtains:

\beq                            \label{310}
\lim_{t \rightarrow 0}u_h^*=\lim_{t \rightarrow 0}
\biggl[\partial_t u_h +\varepsilon \bigl( \dfrac{\lambda^2}{4}u_h - u_{h^{''}}\bigl)\biggr] = h(x).
\eeq

 \section{Solution of the linear  problem}

Let  us consider the homogeneous case. From theorems \ref{t31}, \ref{t32} the following result is obtained:

\bthe  \label{t41} 
 When $ F\,=0\, $ and the initial data  $h_1(x),$  and $   h_0(x)$  verify
the hypotheses of theorems \ref{t31}- \ref{t32}, then the function:

\beq                        \label{41}
u(x,t)=u_{h_1}+(\partial_t+\alpha\,+\,\varepsilon\,\lambda \partial_{x}\,-\varepsilon\partial_{xx})\,u_{h_0}
\eeq

\noindent
 represents a solution of the homogeneous strip problem (\ref{21}).
\ethe

 Otherwise, when  $ F\,=f(x,t)\, $, let consider

\beq                             \label{42}
u_f(x,t)\ \ = \,-\,\int_{0}^{t} d\tau
\int_{0}^{l} f(\xi,\tau) \ \ G(x,\xi,t-\tau)\ \ d\xi.
\eeq

\noindent Standard computations lead to consider at first the problem (\ref{21}) with $ g_0=g_1=0 $. For this    
the following theorem is proved:

\bthe   \label{t42} 
If the function $f(x,t)$ is a continuous function
in $\Omega_T$ with continuous derivative with respect to x, 
 then the function $u_f$
represents a solution of the nonhomogeneous strip problem.

\ethe
{\bf Proof.}
Since $(\ref{33})_1$ it results:

\beq                             \label{43}
\partial_tu_f(x,t)\ \ = \int_{0}^{t} d\tau
\int_{0}^{l} f(\xi,\tau) \ \ G_t(x,\xi,t-\tau)\ \ d\xi
\eeq

\no and as proved in theorem  \ref{t31}, one obtains:

\beq                            \label{44}
\lim_{\tau \rightarrow t}\partial_tu_f(x,t)\, = 
 f(x,t).
\eeq

\noindent
Hence, one has:

\beq                             \label{45}
\partial^2_tu_f\ \ = f(x,t)+
\ \int_{0}^{t} d\tau
\int_{0}^{l} f(\xi,\tau) \ \ G_{tt}(x,\xi,t-\tau)\ \ d\xi
\eeq

\no and theorem \ref{t23} assures that $\,\,{\cal L}_\varepsilon   u_f = f(x,t)\,\, $.

\noindent
Furthermore,
owing to (\ref{42})-(\ref{43}) and estimates (\ref{27}), if  $B_i$ (i=1,2)  are two
positive constants, it results:

\beq                              \label{46}
|u_f|\leq   B_1 (1-e^{-\delta t}); \ \ \ \\ \  |\partial_tu_f|\leq
B_2 (1-e^{-\delta t})\eeq

\noindent
from which initial  homogeneous conditions follow.

\vspace{2mm}The uniqueness is a consequence of the energy-method  and we
have:

\bthe \label{t43} 
When the source term   $f(x,t)$ satisfies
theorem \ref{t42} and the initial data $(h_0,h_1)$  satisfy theorem \ref{t41},
  then the function

\beq                                \label{47}
u(x,t)=u_{h_1}+(\partial_t+\alpha\,+\,\varepsilon\,\lambda \partial_{x}\,-\varepsilon\partial_{xx})\,u_{h_0}+u_f
\eeq
is the unique solution of the linear non-homogeneous
strip problem (\ref{21}). 
\ethe 

\section{Solution of the  non linear  problem}

 \no Let us consider now the non linear problem:

  \beq          \label{51}
  \left \{
   \begin{array}{ll}
    & (\partial_{xx} \, - \,\lambda\,  \partial _x\,)\,\,(\varepsilon
u _{t}+ u) - \partial_t(u_{t}+\alpha\,u)\,= F(x,t, u),\ \  \
       (x,t)\in \Omega_T,\vspace{2mm}\\
   & u(x,0)=h_0(x), \  \    u_t(x,0)=h_1(x), \  \ x\in [0,l],\vspace{2mm}  \\
    & u(0,t)=0, \  \ u(l,t)=0, \  \ 0<t \leq T,
   \end{array}
  \right.
 \eeq

  \no As for the data   $\,F\,  $ and $\,h_i\,(x) \,\,(i=0,1)$  we shall admit:

\vspace{3mm} {\bf Assumption 5.1\,} {\em The functions }$\, h_i(x) \,(i=0,1) $ {\em are  continuously differentiable and bounded together with} $\, h'_1(x) \,$ {\em and }$ h^{(k)}_0 \,\,(k=1,2).$ { \em The function }$\,F(\,x,t,u\,)\, $ {\em is defined and continuous on the set}

\beq  \label{52}
  \,\,\, D_T \ \equiv  \{ \, (x,t,u)\,\,: \,(x,t)  \in   \Omega_T \,, \,\,-\infty \,<\,u\,<\infty \,  \}
 \eeq

\no {\em and more it is  uniformly Lipschitz  continuous in }$ \,( \,x,\,t,\, u\, )\,$ {\em  for  each compact subset of }$\, \Omega_T\,. $ {\em Besides, }$\, F\, $ {\em is bounded for bounded} $\, u\,$ {\em and there exists a constant }$\,C _F\,$ {\em such that the estimate}

\beq \label{53}
\, |F (x,t,u_1)\,-\,F (x,t,u_2)|\, \leq \,\,C _F \,\, \ \, | u_1-u_2\,| \, 
\eeq

 \no {\em holds for all }$\, (\,u_1,\,u_2\,)$. 

 \vspace{2mm}When the problem (\ref{51}) admits a solution $\, u\,$  then, properties of $ G $ and the assumptions 5.1,  assure that $\, u $ must satisfy the integral equation

\beq                                          \label{54}
 u(x,t)=\,
\int_{0}^{l} h_1(\xi) G(x,\xi,t) d\xi
+
(\partial_t+\alpha\,+\,\varepsilon\,\lambda \partial_{x}\,-\varepsilon\partial_{xx})\,\int_{0}^{l} h_0(\xi)
G(x,\xi,t) d\xi +
\eeq
\[+\int_0^ t d\tau\, \int_0^l\, G(x,\xi,t-\tau)\,
F(\xi,\tau,u(\xi,\tau))d\xi,\]

 \no and it is possible to prove that: \cite{c,dmm,dr08} 

\bthe
The non linear problem (\ref{51}) admits a unique solution if and only if the integral equation (\ref{54}) has a unique solution which is continuous on $ \Omega_T $.
\ethe

Moreover,  let  $\,\,|| \, v \, ||_T \,  = \displaystyle{\sup_{\Omega_T}}\,| v (x,t)\,| \,\,$  and let  $ \,{\cal B}_ T \,$  denote the Banach space

\beq     \label{55}
  \,{\cal B}_ T \, \equiv \, \{\, v\,(\,x,t\,) : \, v\, \in  C (\,\Omega_T \,),\,  ||v||_T \, < \infty \ \}.
\eeq

 \no By means of standard methods related to integral equations  it is  possible to prove that the mapping $ \psi $ defined by (\ref{54}) is a contraction of ${\cal B}_T $ in $ {\cal B}_T$ and so it admits  a unique fixed point $ u(x,t).$  In consequence the following theorem holds:

\bthe
When the initial data $h_i\,(i=0,1)$  and the source term $ F $ verify the assumption 5.1, then the  problem (\ref{51}) admits  a  unique regular solution. 
\ethe

 \section{Applications}

All these results allow us to obtain continuous dependence upon the data, a priori estimates of the solution and asymptotic properties.

According to assumption 5.1, let

 \[
\left\| h_i \, \right\| \, =\, \sup_{(0,\,l)} \left|h_i(x) \right|, \,\,\,(i=0,1)\qquad \qquad
\left\| h_0^{''} \, \right\| \, = \sup_{(0,\,l)}\,\, \left|h_0^{''}(x) \right|, \]\,
\[\quad \left\| u \, \right\|_T \, = \sup_{\Omega_T} \left|u(x,t) \right|, \quad \left\| F \, \right\| \, = \sup_{D_T} \left|F(x,t,u) \right|.  
\]

 So, by means of the following theorem the dependence upon the data  can be proved:

\bthe   Let $\,u_1,\,u_2\,$  be two solutions of the problem  related to the data $\, (h_0,\,h_1,\,F_1)\,$  and  
$\, ( \gamma_0,\,\gamma_1\, \,F_2)\,$  which satisfy the assumption 5.1. Then, there exists a positive constant $\,C\,$ such that 

\[
\left\| u_1-u_2 \, \right\|_T \, \leq \, C\, \sup_{\Omega_T }\,\left|h_0-\gamma_0 \right|  \,+\, C\, \sup_{\Omega_T} \,\left|h_1-\gamma_1 \right|  \,+\, C\,  \sup_{D_T} \,\left|F_1(x,t,u) -F_2(x,t,u)\right|,  
\]

\no  where $\,C\, $ depends on $\, C_F, \, T\, $ and on  the parameters $\, \alpha,\,\varepsilon, \lambda.\,$
\ethe

\vspace{3mm}The  integral equation  and  the properties proved for Green Function $ G $ imply a priori estimates, too.

\bthe
When the data $ (h_0,h_1, F)  $ of the  problem (\ref{51}) verify the assumption 5.1, then the following estimate holds:
\beq          
\left\| u(x,t) \, \right\|_T \, \leq \,  \frac{1}{\delta}\,\,\, (1- e^{-\delta\,t} )\,\,\left \| F \right \|  \,+\,  K \,\,[\,\, \left \| h_1 \right \|  \, + \,\, \left \| h_0 \right \| \,\,+ \| h_0^{''}  \|\,\,] \, e^{-\,\delta\,t\,}\, 
\eeq

 \no where the constants $ \delta  $- defined in (\ref{26})- and $ K $ depend on $ \alpha, \varepsilon, \lambda. $
\ethe  

  \vspace{3mm}\no As for the asymptotic properties, obviously the behaviour of the solution depends upon the shape of the source  term.

 For instance, in the linear case one has:

\bthe  When the source term  $f(x,t)$ satisfies the
condition:

\beq                   \label{72}
|\, f(x,t)\,| \, \leq\, C \, e^{-\,m \,t}  \ \\ \ \ \\ \\ \ \ \ \ \ \ \ \
\
(C, \,m = const > 0),
\eeq

\noindent
{\em one has}:

\beq                  \label{73}
\quad |\, u(x,t)\,| \, \leq\, k \, e^{-\,m^* t}  
\ \quad m^* = min\{\delta, \,m\}, \ \ k=const. \eeq

\ethe

\vspace{3mm} An  exponentially decreasing behaviour is also possible  in the non linear case. In fact, according to \cite{ce}, let us  consider  a  normed space where 

\beq    \label{74Ã¢â‚¬Â¢}
\left\| u(x,t) \, \right\| \, = \max_{x \in (0,l)}\,\,|u(x,t) |
\eeq

\no is such that 
\beq      \label{75Ã¢â‚¬Â¢}    
\left\| u(x,t) \, \right\| \, \leq \, \beta  \,\, e^{- \delta \,t}
\eeq

 \no beeing  $ \beta  $ a positive constant and  $ \delta   $  is defined in (\ref{26}). Furthermore, let us introduce   the following definition \cite {ce}:
 
  \vspace{3mm}{\bf Definition 7.1} When the function $ F $ is such that $| F(x,t,u ) \, | \, \leq  \gamma \,||u||\,e^{- \delta\, t}, $ then $ F $ is an exponential Lipschitz function.

   \vspace{3mm}\no So the following theorem can be proved: 
  
\bthe
If the non linear source  $ F  $ is an exponential Lipschitz function, then the solution of the semilinear problem (\ref{51}) vanishes as follows:

\beq
|u(x,t)| \,\leq \,  K_1 \, \, e^{\,-\,\delta\, t}
\eeq 
 
 \no where $ K_1  $ is a positive constant depending on $ \alpha,\,\, \varepsilon,\, \lambda. \,$   
\ethe

 \no Since $\,|\sin \, u | \leq \, |u|, $ a similar  behaviour is also verified   for  the model of superconductivity when $ F(x,t,u)=\,  \sin \, u. \,\,$

\vspace{5mm} \no \textbf{Acknowledgements} This paper has been performed under the auspices of G.N.F.M. of I.N.D.A.M.

\end{document}